\documentclass[11pt]{article}
\usepackage{amsmath,amssymb,color,graphics,epsfig,cite}

\usepackage{amsfonts}

\newcommand{\hoch}[1]{$\, ^{#1}$}

\makeatletter
\@addtoreset{equation}{section}
\makeatother

\newcommand{\be}{\begin{equation}}
\newcommand{\ee}{\end{equation}}
\newcommand{\bea}{\setlength\arraycolsep{2pt} \begin{eqnarray}}
\newcommand{\eea}{\end{eqnarray}}
\newcommand{\nn}{\nonumber}

\def\ft#1#2{{\textstyle{\frac{\scriptstyle #1}{\scriptstyle #2} } }}
\def\fft#1#2{{\frac{#1}{#2}}}

\def\0{{\sst{(0)}}}
\def\1{{\sst{(1)}}}
\def\2{{\sst{(2)}}}
\def\3{{\sst{(3)}}}
\def\4{{\sst{(4)}}}
\def\5{{\sst{(5)}}}
\def\6{{\sst{(6)}}}
\def\7{{\sst{(7)}}}
\def\8{{\sst{(8)}}}
\def\sst#1{{\scriptscriptstyle #1}}

\begin{document}


\begin{center}

{\Large {\bf Bounce Universe and Black Holes from Critical Einsteinian Cubic Gravity}}

\vspace{40pt}
{\bf Xing-Hui Feng, Hyat Huang, Zhan-Feng Mai and  H. L\"u\hoch{*}}

\vspace{10pt}

{\it Center for Advanced Quantum Studies, Department of Physics, \\
Beijing Normal University, Beijing 100875, China}

\vspace{40pt}

\underline{ABSTRACT}
\end{center}

We show that there exists a critical point for the coupling constants in Einsteinian cubic gravity where the linearized equations on the maximally-symmetric vacuum vanish identically.  We construct an exact isotropic bounce universe in the critical theory in four dimensions. The comoving time runs from minus infinity to plus infinity, yielding a smooth universe bouncing between two de Sitter vacua.  In five dimensions we adopt numerical approach to construct a bounce solution, where a singularity occurred before the bounce takes place.  We then construct exact anisotropic bounces that connect two isotropic de Sitter spacetimes with flat spatial sections. We further construct exact AdS black holes in the critical theory in four and five dimensions and obtain an exact AdS wormbrane in four dimensions.

\vfill {\footnotesize \hoch{*}mrhonglu@gmail.com}

\thispagestyle{empty}

\pagebreak

\tableofcontents
\addtocontents{toc}{\protect\setcounter{tocdepth}{2}}



\section{Introduction}

Since the discovery of Einstein's General Relativity, there have been comparable efforts of both studying and modifying the theory.  A natural generalization is to extend the Einstein-Hilbert action with higher-order invariant polynomials of the Riemann tensor.  The generalized theory remains invariant under the general coordinate transformation.  Furthermore, such a higher derivative theory can be renormalizable \cite{Stelle:1976gc,Stelle:1977ry}.  However, when treated on its own, higher-derivative gravities typically suffer from having additional ghostlike massive spin-2 modes in the spectrum.  Recently, new black holes associated with the condensation of the massive spin-2 modes were constructed using numerical approach in Einstein gravity extended with quadratic curvature terms \cite{Lu:2015cqa,Lu:2015psa}.   Interestingly analytical approximate expressions for the metric functions of the black hole in terms of rational polynomials can be constructed using Pad\'e approximants \cite{Kokkotas:2017zwt}.

Alternatively, there exist special combinations of the Riemann tensor polynomials that can render the higher-order theories ghost free, giving rise to Einstein-Gauss-Bonnet gravity or the more general Lovelock gravities \cite{Lovelock:1971yv}.  However, these theories are necessarily in dimensions higher than four.

Recently, Einstein gravity extended with some specific combination of cubic Riemann tensor polynomials were proposed in \cite{Bueno:2016xff}. Like all such higher-order theories, Einsteinian cubic gravity admits maximally-symmetric vacua that are Minkowski, de Sitter (dS) or anti-de Sitter (AdS) spacetimes, depending on the coupling parameters of the theory. The salient feature is that the linearized gravity on these maximally-symmetric vacua is of two derivatives and contains only the graviton modes. This implies that the linearized theory can be ghost free provided that the kinetic term of the graviton is positive.  Thus Einsteinian cubic gravity, which is nontrivial in even four dimensions, is analogous to the Lovelock gravities.  It should be pointed out however ghost excitations associated with higher derivatives in time can still develop in non-maximally symmetric backgrounds in Einsteinian cubic gravity. Although no exact solution of black holes in Einsteinian cubic gravity were known, numerical analysis indicates that a black hole generalizing the Schwarzschild one exists \cite{Hennigar:2016gkm,Bueno:2016lrh} where the black hole properties were also studied.

Einsteinian cubic gravity may in general contain three maximally-symmetric vacua.  For some appropriate choice of the coupling constants, we find that two vacua can coalesce, in which case, the linearized equations of motion become automatically satisfied. This critical phenomenon was also observed in general Lovelock gravities \cite{Crisostomo:2000bb}, and the corresponding critical theory was called ``gravity without graviton'' \cite{Fan:2016zfs}.

One focus of this paper is to construct cosmological solutions in critical Einsteinian cubic gravity.  We consider the standard Friedmann-Lema\^ itre-Robertson-Walker (FLRW) cosmological ansatz which naturally includes the cosmological de Sitter vacuum solution.  Analogous to Einstein gravity or Lovelock gravities, the de Sitter vacuum is rigid in general Einsteinian cubic gravity in that there can be no isotropic scalar perturbation.  However, at the critical point, we find that Einsteinian cubic gravity allows its de Sitter vacuum to deform.  In particular, we obtain an exact solution of an isotropic bounce universe in four dimensions, whose scaling factor is simply $a=\cosh(H_0 t)$, where $t$ is the comoving time.  In five dimensions, an exact solution is lacking and we adopt the numerical approach to establish that a bounce metrics also exists, but the metric contains a branch-cut curvature singularity before the bounce takes place.  We show however that there can be no cosmological isotropic bounces in $D\ge 6$.

We then construct static solutions in critical Einsteinian cubic gravity.  For negative cosmological constant, we obtain exact AdS black holes in both four and five dimensions. We also obtain an exact AdS wormbrane in four dimensions.  When the cosmological constant is positive, we obtain anisotropic bounce universes in four and five dimensions that bounce between two isotropic de Sitter spacetimes.

The paper is organized as follows.  In section 2, we review Einsteinian cubic gravity and derive the critical condition where the linearized equations of motion on a maximally-symmetric vacuum vanishes identically.  In section 3, we study cosmological solutions and obtain isotropic bounce universes in both four and five dimensions.  In section 4, we construct exact static solutions in the critical theories in both four and five dimensions. In section 5, we construct solutions of anisotropic bounce universes. We conclude the paper in section 6.

\section{Critical Einsteinian cubic gravity}

The bulk action of Einsteinian cubic gravity in general $D$ dimensions is given by
\be
S=\fft{\kappa_0}{16\pi}\int d^D x\, \sqrt{-g} L\,,\qquad L= R-2\Lambda_0+\lambda{\cal P}\,,
\ee
where the cubic invariant polynomial term $\cal P$ of the Riemann tensor is \cite{Bueno:2016xff}
\be
{\cal P} = 12R_\mu{}^\rho{}_\nu{}^\sigma R_\rho{}^\gamma{}_\sigma{}^\delta R_\gamma{}^\mu{}_\delta{}^\nu+R_{\mu\nu}^{\rho\sigma}R_{\rho\sigma}^{\gamma\delta}
R_{\gamma\delta}^{\mu\nu}-12R_{\mu\nu\rho\sigma}R^{\mu\rho}R^{\nu\sigma}
+8R_\mu^\nu R_\nu^\rho R_\rho^\mu\,.
\ee
The theory contains a total of three coupling constants: $\kappa_0$ that is related to the inverse of the bare Newton's constant, the bare cosmological constant $\Lambda_0$ and the coupling constant $\lambda$ for the cubic terms. Since we do not consider the case with infinitely-large $\lambda$, we can without loss of generality set $\kappa_0=\pm 1$.

The covariant equation of motion associated with the variation of the metric is \cite{Bueno:2016xff}
\be
{\cal E}_{ab} \equiv \kappa_0 (P_{acde}R_b{}^{cde}-\ft12g_{ab}{L}-2\nabla^c\nabla^dP_{acdb}) = 0\,,
\ee
where
\bea
P_{abcd} &\equiv& \frac{\partial{L}}{\partial R^{abcd}}\nn\\
&=&\ft12(g_{ac}g_{bd}-g_{ad}g_{bc})+6\lambda
\Big(R_{ad}R_{bc}-R_{ac}R_{bd}+g_{bd}R_a{}^eR_{ce}-g_{ad}R_b{}^eR_{ce}\nn\\
&&-g_{bc}R_a{}^eR_{de}+g_{ac}R_b{}^eR_{de}-g_{bd}R^{ef}R_{aecf}
+g_{bc}R^{ef}R_{aedf}+g_{ad}R^{ef}R_{becf}\nn\\
&&-3R_a{}^e{}_d{}^fR_{becf}-g_{ac}R^{ef}R_{bedf}+3R_a{}^e{}_c{}^fR_{bedf}
+\ft12R_{ab}{}^{ef}R_{cdef}\Big)\,.\label{pabcd}
\eea
The vacua of the theory are maximally-symmetric spacetimes with the Riemann tensor
\be
\bar R_{abcd} = \frac{2\Lambda_{\rm eff}}{(D-1)(D-2)}(\bar g_{ac}\bar g_{bd}-\bar g_{ad}\bar g_{bc})\,.
\ee
The effective cosmological constant $\Lambda_{\rm eff}$ satisfies the cubic order algebraic equation \cite{Bueno:2016xff}:
\be
-\fft{4\tilde \lambda}{27\Lambda_0^2} \Lambda_{\rm eff}^3+\Lambda_{\rm eff}-\Lambda_0 = 0\,,
\label{lameffeq1}
\ee
where we introduce a dimensionless coupling constant
\be
\tilde \lambda\equiv \frac{108(D-3)(D-6)}{[(D-1)(D-2)]^2}\lambda\Lambda_0^2
\,.\label{lameffeq2}
\ee
Positive, zero or negative $\Lambda_{\rm eff}$'s yield dS, Minkowski or AdS vacua respectively.  The Minkowski spacetime arises only when $\Lambda_0=0$, and asymptotically flat black holes were constructed and studied in \cite{Hennigar:2016gkm,Bueno:2016lrh}.  In this paper, we consider only the case with non-vanishing $\Lambda_0$.

\begin{figure}[htp]
\begin{center}
\includegraphics[width=180pt]{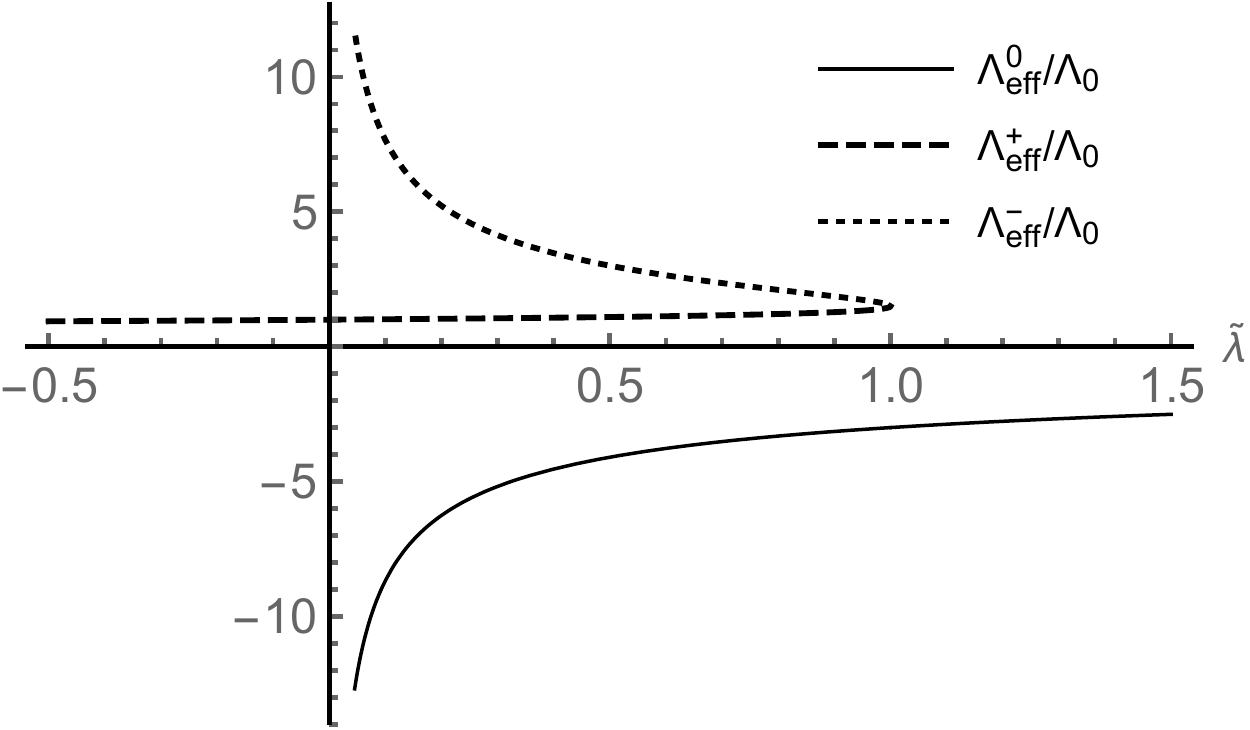}\ \ \
\includegraphics[width=180pt]{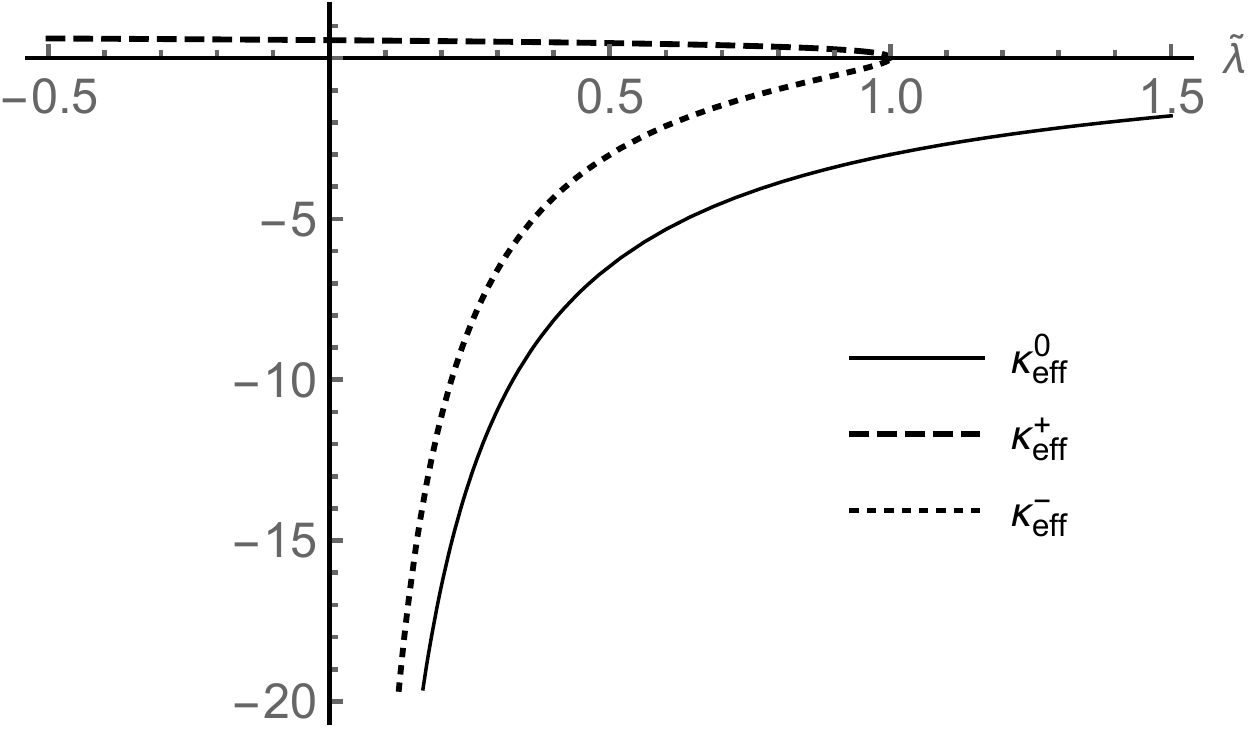}
\end{center}
\caption{\small\it The left figure plots the effective cosmological constant $\Lambda_{\rm eff}$ of the vacua.  There is at least one (A)dS vacuum for all $\tilde\lambda$. In the region $0<\tilde \lambda<1$, there are three (A)dS vacua.  The dashed and dotted vacua coalesce to become one vacuum when $\tilde \lambda=1$.  The right figure plots the effective $\kappa_{\rm eff}$ of the linearized graviton in each vacuum, with $\kappa_0=1$.  We see that ghost free vacuum (dashed-line) exists only for $\tilde\lambda\le 1$, where $\kappa_{\rm eff}\ge 0$, with $\kappa_{\rm eff}=0$ at $\tilde \lambda=1$.  For $\kappa_0=-1$, The vacua associated with dotted and solid lines becomes ghost free instead.} \label{LeffKeff}
\end{figure}

In three and six dimensions, we have $\tilde \lambda=0$ and hence there exists only one effective cosmological constant, which is the same as the bare one. In general dimensions, there can be three roots to (\ref{lameffeq1}), given by
\bea
\Lambda^0_{\rm eff} &=& -\fft{3\Lambda_0(1 + (\sqrt{\tilde\lambda} + \sqrt{\tilde \lambda-1})^{2/3})}{2\sqrt{\tilde\lambda} (\sqrt{\tilde\lambda} + \sqrt{\tilde \lambda-1})^{1/3}}\,,\nn\\
\Lambda^\pm_{\rm eff} &=& \fft{3\Lambda_0\tilde\lambda(1\mp {\rm i}\sqrt3 +
(1 \pm {\rm i}\sqrt3) (\sqrt{\tilde\lambda} + \sqrt{\tilde\lambda-1})^{2/3})
}{4\tilde\lambda^{3/2}(\sqrt{\tilde\lambda} + \sqrt{\tilde \lambda-1})^{1/3}}\,.\label{3lameff}
\eea
Thus for $\tilde\lambda\le 0$ and $\tilde\lambda>1$, there is only one real root, given by
$\Lambda^+_{\rm eff}$ and $\Lambda^0_{\rm eff}$ respectively, and the remaining two roots form a complex conjugate pair.  For $0<\tilde \lambda\le 1$, all three roots are real.  In particular, when $\tilde\lambda=1$, the $\Lambda_{\rm eff}^+$ and $\Lambda_{\rm eff}^-$ becomes the same, namely
\be
\tilde\lambda=1\qquad \rightarrow\qquad \Lambda^+_{\rm eff}=\Lambda^-_{\rm eff}=\Lambda^*_{\rm eff}\equiv\ft32\Lambda_0\,,\qquad\Lambda_{\rm eff}^0=-3\Lambda_0\,.\label{critlamb}
\ee
Thus at the critical point of the parameter $\lambda=\lambda^{\rm cr}$, where
\be
\lambda^{\rm cr} \Lambda_0^2= \frac{(D-1)^2(D-2)^2}{108(D-3)(D-6)}\,,\label{critpoint}
\ee
the two vacua of $\Lambda^\pm_{\rm eff}$ coalesce to become one with the effective cosmological constant $\Lambda_{\rm eff}^*=\ft32\Lambda_0$.  The quantities $\Lambda^0_{\rm eff}$ and $\Lambda_{\rm eff}^\pm$ when they are real are plotted in the left figure of Fig.~\ref{LeffKeff}.

Having analysed the vacua of the theory, we now consider the linear perturbation of these vacua:
\be
g_{ab} = \bar g_{ab}+h_{ab}\,.
\ee
The linearized equation of motion for $h_{ab}$ is given by \cite{Bueno:2016xff}
\be
\kappa_{\rm eff}\, G^{\rm L}_{ab}=0\,\qquad\hbox{with}\qquad
\kappa_{\rm eff}=\kappa_0\left(1-\frac{48(D-3)(D-6)\lambda\Lambda_{\rm eff}^2}{(D-1)^(D-2)^2}\right)\,,
\ee
where
$\kappa_0 G^{\rm L}_{ab}=0$ is the linearized equation of Einstein gravity on the (A)dS vacuum of $\Lambda_{\rm eff}$, with
\bea
G^{\rm L}_{ab} &=& -\frac{1}{2}\Big(\frac{4\Lambda_{\rm eff}h_{ab}}{D-2}-\frac{2\Lambda_{\rm eff}\bar g_{ab}h}{D-2}+\nabla_b\nabla_ah-\nabla_c\nabla_ah_b{}^c-\nabla_c\nabla_bh_a{}^c\nn\\
&&\qquad +\nabla_c\nabla^ch_{ab}+\bar g_{ab}\nabla_d\nabla_ch^{cd}-\bar g_{ab}\nabla_c\nabla^ch\Big)\,.
\eea
Thus if $\kappa_{\rm eff}<0$, the corresponding graviton mode $h_{ab}$ has negative kinetic energy and is ghostlike.

As can be seen in the right plot of Fig.~\ref{LeffKeff}, when $\kappa_0=1$, only the $\Lambda^+_{\rm eff}$ vacuum, arising from $\tilde\lambda<1$, have positive kinetic energies for the linear graviton modes.   When $\kappa_0=-1$, on the other hand, the $\Lambda_{\rm eff}^0$ and $\Lambda_{\rm eff}^-$ vacua are ghost free.  Analogous properties in Einstein-Gauss-Bonnet gravity were studied \cite{Canfora:2013xsa}. In fact the behavior of the two branches $\Lambda^\pm_{\rm eff}$ has the same qualitative features of the two vacua in Einstein-Gauss-Bonnet gravity. The kinetic term vanishes identically at the critical point $\tilde \lambda^{\rm ct}=1$, giving rise to ``gravity without graviton'' \cite{Fan:2016zfs}.  In this case, the sign choice of $\kappa_0$ is immaterial.

In the following sections, we shall study the cosmology and black holes in critical Einsteinian cubic gravities.

\section{Isotropic Bounce universes in $D=4,5$}

\subsection{Cosmology in general dimensions}

We begin with the standard FLRW cosmological ansatz in general $D$ dimensions
\be
ds_D^2=-dt^2 + a(t)^2 dx^i dx^i\,,\qquad i=1,2,\cdots,D-1\,.
\ee
The metric is both homogeneous and isotropic. For both simplicity and relevance with cosmological observations, we focus only flat spatial directions.  Furthermore, we shall consider pure Einsteinian cubic gravity without including any matter.  The equations of motion reduces to the following 3'rd-order nonlinear differential equation
\be
\dddot a = \fft{U}{V}\,,\label{dddota}
\ee
where a dot denotes a derivative with respect to $t$ and
\bea
U&=& 2 \Lambda_0 a^6 -(D-1) (D-2) a^4 \dot a^2 +4(D-1)(D-2)(D-3)\lambda\Big(
4a^3 \ddot a^3\nn\\
&&\qquad\qquad -6(D-5)a^2 \dot a^2 \ddot a^2 + 12 (D-5)a \dot a^4 \ddot a - 5(D-4)\dot a^6\Big)
\,,\nn\\
V&=& 48(D-1) (D-2) (D-3) \lambda a^2 \dot a(a \ddot a-\dot a^2)\,.
\eea
We now assume that the theory admits the dS spacetime as a cosmological solution with
\be
a=a_0\, e^{H_0 t}\,,
\ee
where $H_0$ is the Hubble constant.  The equation of motion requires that
\be
\Lambda_0 = \ft12 (D-1)(D-2) H_0^2 \Big(1 - 4(D-3)(D-6) H_0^4\lambda\Big)\,.\label{genLam}
\ee
We are interested in constructing an FLRW solution that is not the de Sitter spacetime but is asymptotic to the dS at the infinite future.  The existence of such a solution can be established by considering an isotropic scalar perturbation, namely
\be
a=a_0\, e^{H_0t} (1 + \tilde a)\,.
\ee
At the linear level, we find
\be
\Big(1 - 12 (D-3)(D-6)H_0^4\lambda\Big)\dot{ \tilde a}=0\,.
\ee
Thus for the generic parameter $\lambda$, we have $\dot{\tilde a}=0$ and hence the de Sitter vacuum is rigid and cannot be deformed by this perturbation. However, when we have
\be
\lambda = \fft{1}{12(D-3)(D-6)H_0^4}\,,\label{genlam}
\ee
we can have non-vanishing $\dot{\tilde a}$, since the linear equation vanishes identically.  It turns out that equations (\ref{genlam}) and (\ref{genLam}) give rise to precisely the critical point described in the previous section and the resulting effective cosmological constant $\Lambda_{\rm eff}=\ft12 (D-1)(D-2)H_0^2$ is $\Lambda_{\rm eff}^*$ defined in (\ref{critlamb}).

Thus we see that the de Sitter vacuum is in general rigid, but it can be deformed isotropically at the critical point.  At the quadratic order, the equation for $\tilde a$ becomes
\be
2\ddot {\tilde a} \dddot {\tilde a} + (D-1) H_0 \ddot {\tilde a}^2 - 2(D-6) H_0^3 \dot {\tilde a}^2=0\,.
\ee
We find a solution, given by
\be
\tilde a=e^{-\nu H_0 t}\,,\qquad 2\nu^3 - (D-1)\nu^2 + 2(D-6)=0\,.
\ee
In $D=4,5$, the cubic polynomial equation for $\nu$ has only one real root, which is positive. For $D\ge 7$ dimensions, all three roots are real, two of which are positive.  For a given positive $\nu$, we can perform large $t$ asymptotic expansion
\be
a = a_0 e^{H_0t}\Big(1 + c_1 e^{-\nu H_0 t} + c_2 e^{-2\nu H_0 t} + c_3 e^{-3\nu H_0 t} +
\cdots\Big)\,.
\ee
We find that the constants $c_2, c_3, etc.$~can be determined in terms of $c_1$ by the equation (\ref{dddota}) order by order of $e^{-\nu H_0 t}$. This shows that there exist cosmological solutions that are asymptotic to the de Sitter spacetime in critical Einsteinian cubic gravity.  It is worth pointing out that there is no such solution in Einstein-Gauss-Bonnet gravity and the de Sitter vacuum of Einstein-Gauss-Bonnet gravity remains rigid even at the critical point.

In this paper, we are interested in cosmology where a bounce occurred before the de Sitter inflation.  Without loss of generality, we assume that the bounce occurs at $t=0$, which implies $\dot a(0)=0$ and $\ddot a(0)>0$. Such a bounce universe cannot arise in classical theory of Einstein gravity since it will violate the null energy condition.  This can be seen easily that in Einstein gravity, the matter energy and pressure densities $\rho$ and $p$ satisfy
\be
\rho + p = (D-2)\Big(\fft{\dot a^2}{a^2} - \fft{\ddot a}{a}\Big)\,,
\ee
which is necessarily negative at the bounce point, violating the null energy condition.
However, such bounce solutions can arise in suitable higher-derivative gravities, since the Einstein tensor is no longer directly related to the matter energy-momentum tensor.  Around the bounce point $t=0$, we can perform Taylor expansion
\be
a=a_{\rm min} (1 + a_2 t^2 + a_3 t^3 + a_4 t^4 + \cdots)\,.\label{bouncet=0}
\ee
The bouncing is symmetric if the odd-power terms vanish.  Substituting the above ansatz into the equation and solve it order by order for small $t$, we find
\be
a_2^3 = \ft{1}{16}(6-D) H_0^6\,.
\ee
Thus the coefficient $a_2$ can only be positive in $D=4$ and $D=5$ dimensions.  In the next two subsections, we shall construct the asymptotically-dS bounce solutions in these dimensions.

\subsection{Bounce universe in four dimensions}
\label{d4bounce}

In four dimensions, the critical point is given by
\be
\lambda^{\rm ct} \Lambda_0^2=-\ft16\,,\qquad \hbox{with}\qquad \Lambda_{\rm eff}^* = \ft32\Lambda_0=3H_0^2\,.\label{d4h0}
\ee
The bounce universe is particularly simple, given by
\be
a=a_0 \cosh(H_0 t)\,.\label{d4a}
\ee
This is a classic bounce solution.  We have chosen the integration constant such that the bounce occurs at $t=0$. As the comoving time runs from $-\infty$ to $+\infty$, the universe bounces between two asymptotic dS vacua of the same cosmological constant.  Note that the solution is valid even with the spatial metric $dx^i dx^i$ replaced by those of $S^2$ or $H^2$.

  It is of interest to examine the stability of the solution against isotropic scalar perturbation. We consider
\be
a=a_0 \cosh(H_0 t) (1 + \phi(t))\,.
\ee
At the linear order, we find
\be
\cosh(H_0 t) \sinh(2H_0 t) \dddot \phi + H_0 \cosh(H_0t) (\cosh(2H_0t)-3) \ddot \phi -
4 H_0^2 \sinh^3(H_0 t)\dot \phi=0\,.
\ee
The linear equation can be solved explicitly, given by
\be
\phi= \phi_0 + \phi_1 \tanh(H_0 t) +\phi_2 \sinh(H_0t)\,.
\ee
The first two modes are convergent, corresponding to constant scaling of $x^i$ and constant shifting of $t$ respectively and hence they do not alter the property of the solution.  The third mode is divergent, indicating instability of the bounce solution.  For small $t$, the $\phi_2$ mode has odd powers of $t$ in the Taylor expansion.  This implies that for the bouncing (\ref{bouncet=0}) with asymmetric terms of odd powers, the solution describes a bouncing universe between two singularities. The bouncing between two asymptotic de Sitter spacetimes as in (\ref{d4a}) thus requires fine tuning.

\subsection{Bounce universe in five dimensions}

The critical point in five dimensions is
\be
\lambda^{\rm ct} \Lambda_0^2 = -\ft23\,,\qquad \hbox{with}\qquad \Lambda_{\rm eff}^* = \ft32\Lambda_0=6H_0^2\,.\label{d5h0}
\ee
Asymptotically at large $t$, we have
\be
a_{\rm asym}=a_0 e^{H_0t} \Big(1 + \mu e^{-\nu H_0 t} + c_2 \mu^2 e^{-2\nu H_0 t} +
c_3 \mu^3 e^{-3 \nu H_0 t} + \cdots\Big)\,,\label{d5asym}
\ee
where $\nu\sim 2.2056$ is the real root of the cubic polynomial $\nu^3-2\nu^2-1=0$, and
\bea
c_2 &=& -\frac{4 \nu ^2+3 \nu -4}{24 \left(\nu ^2+1\right)}\sim -0.1568\,,\nn\\
c_3 &=& \frac{12533 \nu ^2+2476 \nu +5720}{864 \left(\nu ^2+1\right)^2 \left(6 \nu ^2+5\right)}\sim 0.07102\,.
\eea
Note the parameter $\mu$ represents the freedom of shifting the comoving time coordinate $t$.  We now construct the bounce solution whose asymptotic structure is given  by (\ref{d5asym}).  Assuming that such a solution exists, we can choose $\mu$ appropriately so that the bounce occurs precisely at $t=0$, for which the Taylor expansion reads
\bea
a_{\rm bounce} &=& a_{\rm min} \Big(1 + \ft{H_0^2}{(16)^{1/3}}\, t^2 + a_3 t^3 + a_4 t^4 + \cdots\Big)\,,\nn\\
a_4 &=&  \ft1{48} (1 - 2^{1/3}) H_0^4 - \ft{3a_3^2}{2^{3/2} H_0^2}\,.
\eea
Thus at the bouncing point, the equation allows to have an extra free parameter $a_3$. A fine tuning is necessary to choose a specific $a_3$ so that the solution can integrate out to asymptotic infinity.  For a generic $a_3$, the solution describes a bounce between two spacetime singularities.

\begin{figure}[htp]
\begin{center}
\includegraphics[width=180pt]{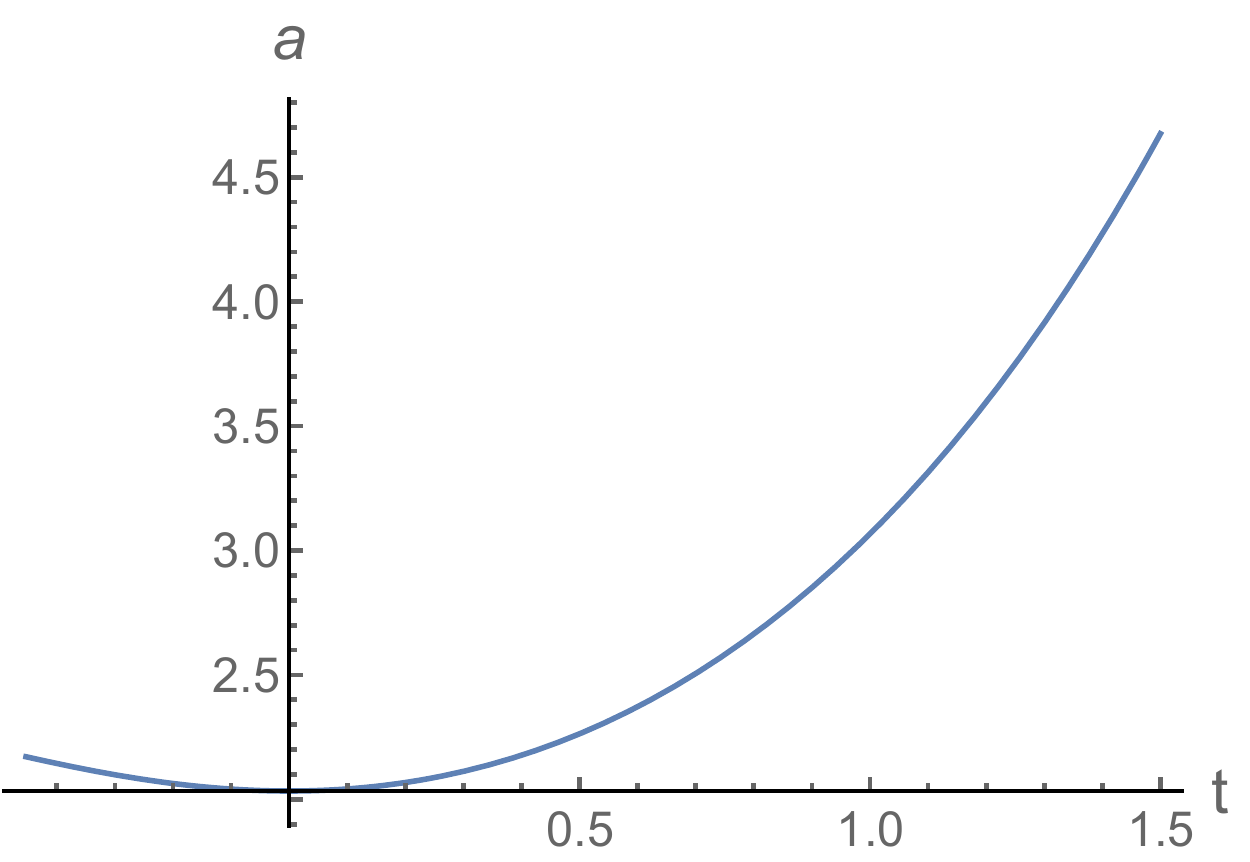}\ \ \
\includegraphics[width=180pt]{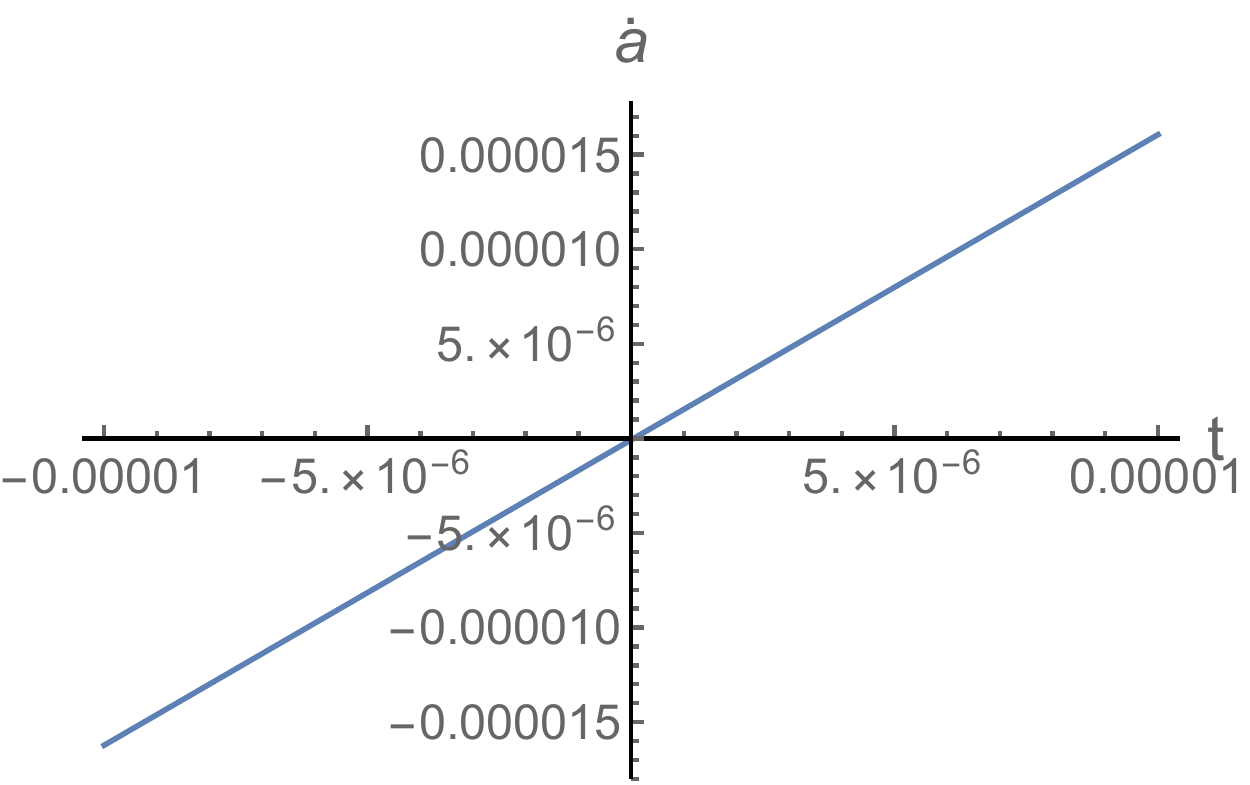}
\end{center}
\caption{\small\it The left is the scaling factor $a$ as a function of comoving time. The right is $\dot a$ near $t=0$.  We choose the parameter $\mu\sim 1.1861$ so that the bounce occurs at $t=0$, and it reaches asymptotic de Sitter as $t\rightarrow \infty$.  The solution has a naked singularity in the past at $t_*=-0.4531$, before the bounce takes place at $t=0$.} \label{d5a(t)}
\end{figure}

We do not find an exact solution of the bounce universe. We can however establish that such a solution indeed exists numerically.  In practice, we find it is more convenient to use the asymptotic expansion (\ref{d5asym}), which we expand to include the seventh order, as our initial data.  Without loss of generality we set $H_0=1$ and $a_0=1$.  We find that if we set $\mu\sim 1.1861$, the bounce occurs at $t\sim 0$.  The results were plotted in Fig.~\ref{d5a(t)}.

\begin{figure}[htp]
\begin{center}
\includegraphics[width=180pt]{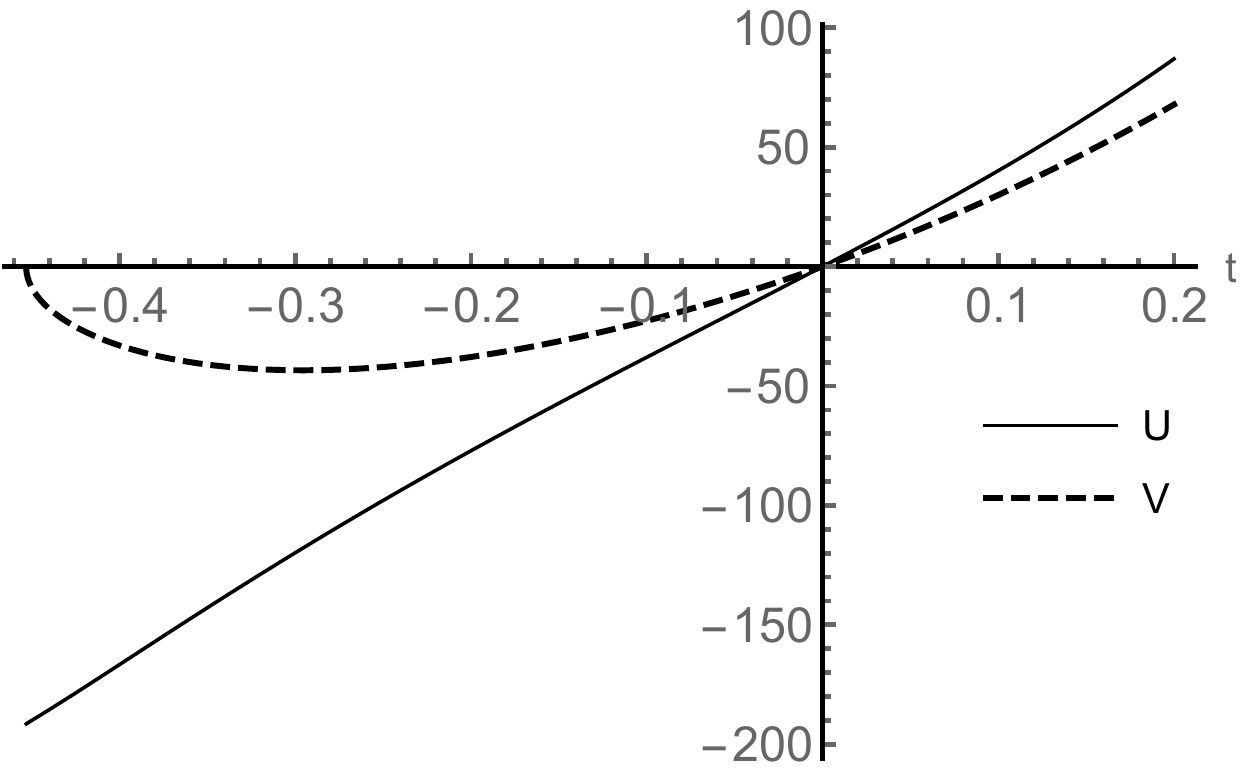}\ \ \
\end{center}
\caption{\small\it The plots for $U$ and $V$, the numerator and denominator in (\ref{dddota}) respectively. $U$ and $V$ both vanishes at the bounce time $t=0$, giving rise to finite $\dddot a$. At $t=t_*$, only $V$ vanishes but not $U$, and hence $\dddot a$ is divergent.} \label{uvplot}
\end{figure}

Our numerical analysis indicates that the minimum scale of the bounce universe is $a_{\rm min}=2.0336$.  Near the bounce point $t\sim 0$, the scaling factor $a$ behaves as
\be
a_{\rm bounce}=2.0336\Big(1 + 0.3969 t^2 + 0.1197 t^2 + \cdots\Big)\,,\label{d5abounce}
\ee
Thus, we see that the coefficient $a_3$ is fixed at $a_3=0.1197$.  Unlike the smooth $D=4$ bounce solution, the universe has a pre-bounce singularity at $t_*\equiv -0.4531$. The origin of this singularity is not that the Riemann tensor diverges at $t=t_*$, but rather the denominator $V$ in (\ref{dddota}) vanishes at $t=t_*$, as can be seen in Fig.~\ref{uvplot}. Such singularity is unlikely to exist in Einstein gravity, but not uncommon in higher-derivative gravities, where Riemann tensors at the singularity are regular whilst the covariant derivatives of the Riemann tensor are singular.

\begin{figure}[htp]
\begin{center}
\includegraphics[width=120pt]{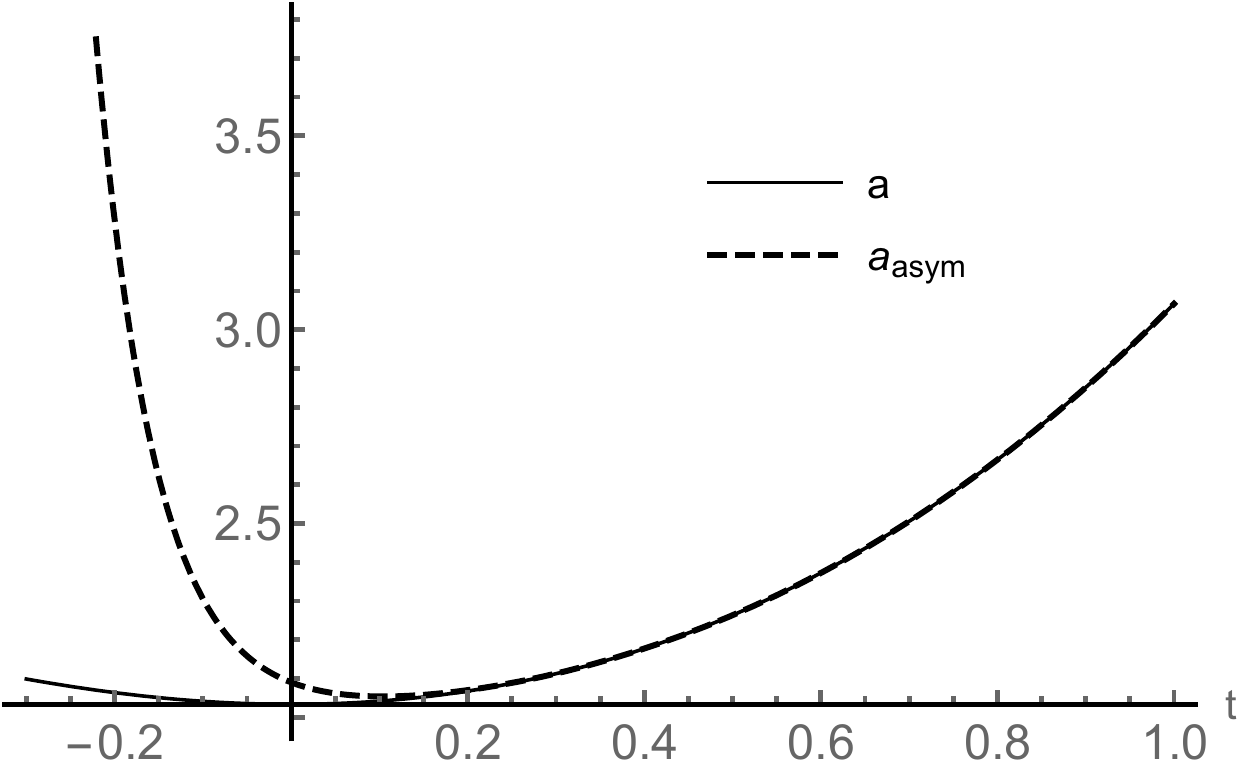}\ \ \
\includegraphics[width=120pt]{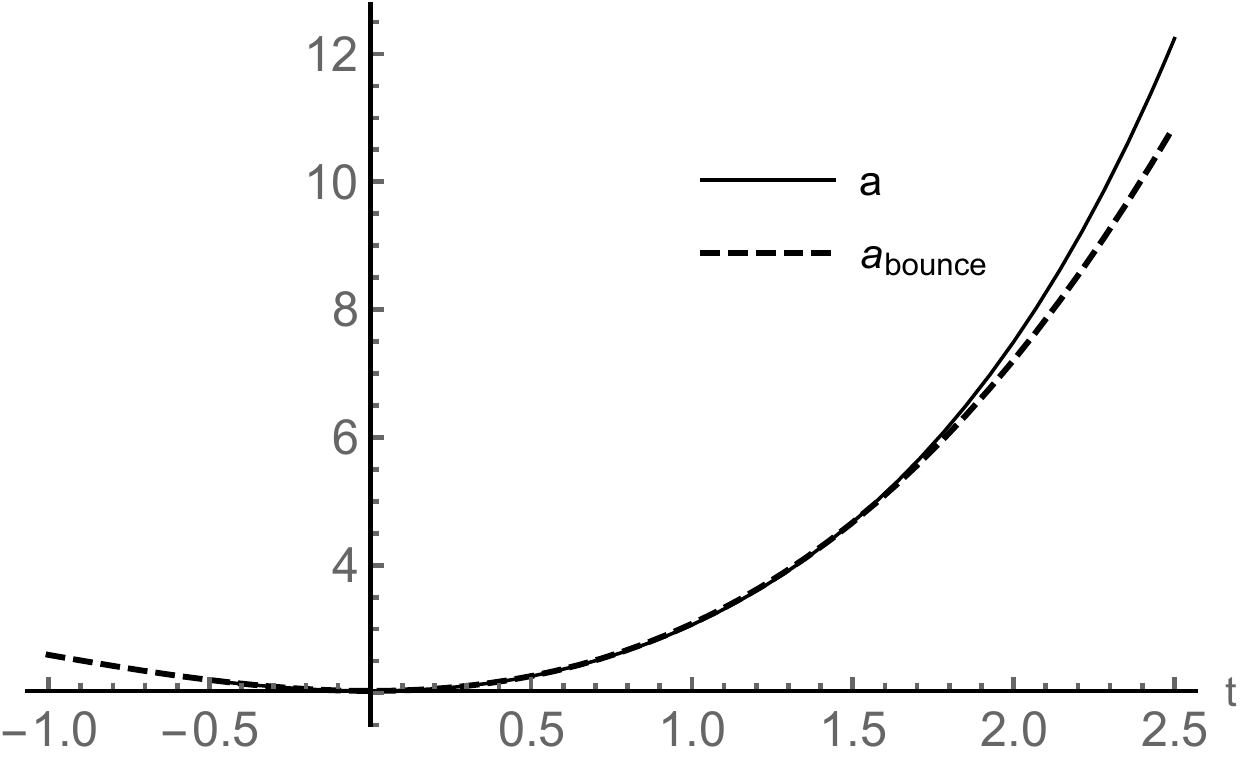}\ \ \
\includegraphics[width=120pt]{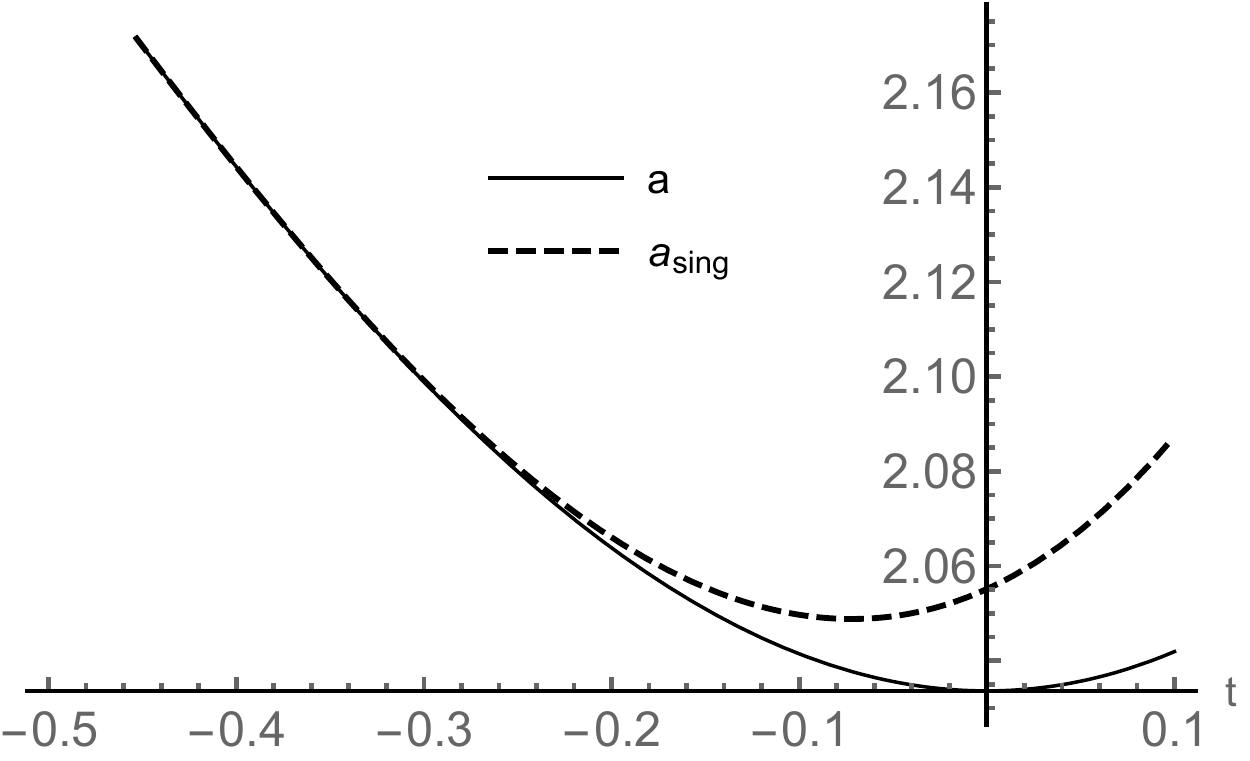}\ \ \
\end{center}
\caption{\small\it Here we plot the numerical $a$ and approximation at three different comoving time regions.  The solid line represents the numerical $a$ and the dashed lines are approximated solutions.  The left plot is at large $t$ region with $a_{\rm symp}$ given in (\ref{d5asym}); the middle plot is at the bounce $t=0$ region with $a_{\rm bounce}$ given in (\ref{d5abounce}); the right plot is at the singular $t=t_*$ region with $a_{\rm sing}$ given in (\ref{d5asing}).} \label{approx}
\end{figure}

By numerical approach, we find that the scaling factor $a$ near the singularity $t=t_*$ can be approximated by the function
\be
a_{\rm sing} \sim  1.9476 - 0.4667 t + 0.0600 t^2 +
(t-t_*)^{3-\ft14\nu} \Big( 0.5413 + 0.4547 (t-t_*)\Big)\,.\label{d5asing}
\ee
It can be easily seen that $\dddot a_{\rm sing}\sim (t-t_*)^{-\nu/4}$ is singularity at $t=t_*$.  In Fig.~\ref{approx}, we plot the the numerical result of $a(t)$ and approximated solutions $a_{\rm asym}$, $a_{\rm bounce}$ and $a_{\rm sing}$ at large $t$, $t=0$ and $t=t_*$ regions.   The solution demonstrate that in higher derivative theories, bounce universe can have singularity at some past time before the bounce takes place.  The vast different behaviors of the cosmological solutions in different dimensions demonstrate that Einsteinian cubic gravities are rather different from the Einstein-Gauss-Bonnet or Lovelock series, where the dimension parameter typically does not alter the characteristics of the solutions in any significant way.

\section{Static solutions in the critical theory}

\subsection{$D=4$ AdS black holes}

\subsubsection{Equations of motion}

We begin with four dimensions.  The metric ansatz for the $D=4$ black holes is given by
\be
ds^2= -f dt^2 + \fft{dr^2}{f} + r^2 d\Omega_{2,k}^2\,,
\ee
where $d\Omega_{2,k}^2$ with $k=1,0,-1$ denotes the metrics for maximally-symmetric spaces of $S^2$, $T^2$ and $H^2$ respectively.  The equations of motion reduces to a 3'rd-order nonlinear differential equation \cite{Hennigar:2016gkm}
\bea
0=&&\ft12(r f' + f -k +\Lambda_0 r^2) +\lambda
\Big(-3 f f''^2+\frac{12 f (f-k) f'}{r^3}+\frac{3 (k-4 f) f'^2}{r^2}\nn\\
&&+\big(\frac{6 f (f-k)}{r}-3 f f'\big) f^{(3)} + \big(\frac{12 f f'}{r}+\frac{12 f (k-f)}{r^2}\big)f'' \Big)\,,\label{d4bheom}
\eea
where a prime denotes a derivative with respect to $r$.  The equation can be integrated, giving \cite{Bueno:2016lrh}
\be
2m=-r(f + \ft13\Lambda_0 r^2 -k) +\lambda\Big(
\frac{12 f (k-f) f'}{r^2}+\frac{6 k f'^2}{r}+2 f'^3+\big(\frac{12 f (f-k)}{r}-6 f f'\big)
f'' \Big)\,.\label{d4fo}
\ee
It turns out that the integration constant $m$ is related to the mass of the black hole.  In fact, $m$ is precisely the mass of the Schwarzschild black hole when $\lambda=0$.

Although there is no exact solution to the equation (\ref{d4fo}), many properties of the black holes can be extracted owing to the fact that $m$ can be identified as the mass \cite{Bueno:2016lrh}.  In this section, we are interested in solutions at the critical point (\ref{critpoint}).  We now consider the case with negative bare cosmological constant $\Lambda_0$, and we parameterize the effective cosmological constant $\Lambda_{\rm eff}^*=-3g^2$ where $1/g$ is the radius of the AdS vacuum.  Thus we have
\be
\Lambda_0=-2g^2\,,\qquad \lambda = -\fft{1}{24g^4}\,.\label{d4crit}
\ee

\subsubsection{Exact black hole as the thermal vacuum}

We find that at the critical point (\ref{d4crit}), there exists an exact solution
\be
f=g^2 r^2 + k - \mu\,.\label{thermalvac}
\ee
It describes a black hole that is asymptotic to AdS with a horizon at $r_0$ with
\be
\mu=g^2 r_0^2 + k\,.\label{mur0}
\ee
The temperature of the black hole is given by
\be
T=\fft{g^2r_0}{2\pi}\,.
\ee
From the Wald entropy formula \cite{wald1,wald2}, we find that the entropy of these static black holes in general dimensions are given by
\be
S=\ft14 \omega r_0^{D-2} \Big( 1 + \fft{3(D-2)(D-3)\lambda f'(r_0)}{r_0^3} \big( r_0 f'(r_0) + 4 k\big)\Big)\,.\label{genentropy}
\ee
For the solution (\ref{thermalvac}), the entropy is a pure constant, independent of $r_0$, namely
\be
S=-\fft{\kappa_0 \omega}{2g^2} k\,.
\ee
This implies that the mass of the black hole must vanish.  In fact, substituting the solution (\ref{thermalvac}) to (\ref{d4fo}) shows that the mass parameter $m$ indeed vanishes identically. As we shall see in the next subsection, the mass of this solution derived from the Wald formalism indeed vanishes also.

Analogous solutions with non-zero temperature, but vanishing mass and entropy (or constant entropy) were also found in critical gravity and conformal gravity, and were referred as black hole thermal vacua \cite{Lu:2011zk,Lu:2012xu}.

\subsubsection{Wald formalism}
\label{wald}

The Wald formalism \cite{wald1,wald2} is a very useful tool to establish the first law of black hole thermodynamics. It can also be used to evaluate the mass of a solution by computing the variation of the Hamiltonian associated with the time-like Killing vector $\xi$, namely
\be
\delta {\cal H} = \fft{\kappa_0}{16\pi} \int \delta {\bf Q} - {i}_\xi {\bf \Theta}\,.\label{varyH}
\ee
Here $\bf Q$ is a $(D-2)$-form that is the Hodge dual of the 2-form with the components
\be
Q^{ab} = 2P^{abcd} \nabla_c \xi_d - 4 \xi_d \nabla_c P^{abcd}\,,
\ee
where $P^{abcd}$ is given by (\ref{pabcd}), and $\bf \Theta$ is a $(D-1)$-form that is the Hodge dual to the 1-form
\be
J^a=2P^{abcd} \nabla_d \delta g_{bc} - 2\delta g_{bc} \nabla_d P^{abcd}\,.
\ee
Note that all the variations are performed on the integration constants of the solutions. For a black hole, the first law of thermodynamics can be derived from the identity
\be
(\delta {\cal H})_{r=r_0}=(\delta {\cal H})_{r\rightarrow\infty}\,.\label{waldid}
\ee

Substituting the black hole thermal vacuum solution in the previous subsection into (\ref{varyH}) and evaluate at the asymptotic infinity, we find that $(\delta {\cal H})_{r\rightarrow \infty}$ vanishes, indicating that the mass of the solution is zero.  We find also that $(\delta {\cal H})_{r=r_0}=0$, and thus the first law of thermodynamics is trivially satisfied.
(Wald formalism for quadratic extended gravity and general Riemann tensor polynomial gravities can be found in \cite{Fan:2014ala} and \cite{Bueno:2016ypa} respectively.)

\subsubsection{Small mass black hole solution}

In the previous subsection, we obtain an exact solution describing a black hole thermal vacuum with vanishing mass.  We now consider small perturbation to include the mass parameter $m$. At the linear order of $m$, we find
\bea
f &=& g^2(r^2-r_0^2) \Big(1 - m \tilde f\Big)\,,\nn\\
\tilde f &=&\fft{1}{g^2 r_0^2 +k}\Big(\fft{r^2 + r_0 r+ 2r_0^2}{2r_0^2(r+r_0)} -
\fft{r^2-r_0^2}{2r_0^3} {\rm arctanh}\big(\fft{r_0}{r}\big)\Big)\,.\label{smallmasssol}
\eea
Here we expressed $\mu$ in terms of the horizon $r_0$ as in (\ref{mur0}). Note that the function $\tilde f$ is governed by a third-order differential equation at the linear order, and can be solved exactly.  We have fixed the three integration constants so that $\tilde f$ presented above is convergent at both $r=r_0$ and $r=\infty$. In other words, the solution satisfies the equations (\ref{d4bheom}) or equivalently (\ref{d4fo}) at the linear order of $m$. Since the function $\tilde f$ is monotonically decreasing from finite positive value at the horizon $r_0$ to zero at asymptotic infinity, it follows that (\ref{smallmasssol}) is a good solution provided that
\be
0\le m \ll \fft{1}{\tilde f(r_0)}= r_0(g^2 r_0^2 + k)\,.
\ee
Asymptotically, the metric function $f$ reads
\be
f=g^2(r^2 -r_0^2) + \fft{g^2 m}{g^2 r_0^2 + k} \Big(-\ft43 r + r_0 + \fft{4r_0^2}{15r} +
\fft{4r_0^4}{105r^3} + \cdots \Big)\,.
\ee
Substituting this into (\ref{varyH}), we find that
\be
(\delta {\cal H})_{r\rightarrow \infty} = \fft{\kappa_0\omega}{4\pi} \delta m\,,
\ee
which implies that the mass of the solution is
\be
M=\fft{\kappa_0 \omega}{4\pi} m\,.
\ee
The horizon is still located at $r=r_0$, but the near horizon geometry is not of the $\mathbb{R}^2\times S^2$, but with $\mathbb{R}^2$ replaced by a two-dimensional space with curvature singularity.  To see this, we note that near the horizon, we have
\be
f=2g^2 \big(r_0 - \fft{m}{g^2 r_0^2 + k}\big)(r-r_0) - \fft{g^2 m}{r_0 (g^2r_0^2 + k)}
 (r-r_0)^2 \log\big(\fft{r-r_0}{2r_0}\big) + {\cal O}((r-r_0)^2)\,.
\ee
The logarithmic term above implies that although we have $f(r_0)=0$ and finite $f'(r_0)$, the quantity $f''(r_0)$ is divergent.  In fact, the Ricci scalar of the solution is given by
\bea
R&=&
\frac{g^2 m \left(15 r^3 r_0+r r_0^3-2 r_0^4 -\left(15 r^4-12 r_0^2 r^2+r_0^4\right) {\rm arctanh}\left(\frac{r_0}{r}\right)\right)}{r^2 r_0^3 \left(g^2 r_0^2+k\right)}\nn\\
&&+ \fft{2(g^2 (r_0^2-6r^2)+k)}{r^2} + {\cal O}(m^2)\,,
\eea
which is divergent at $r=r_0$.  However, the fact that curvature singularity coincides with the horizon implies that the singularity is not naked, and the solution can be viewed as a black hole.

Regardless the divergence of $f''$ on the horizon, we may still impose the vanishing of conic singularity associated with the Euclidean time and obtain the temperature
\be
T=\fft{f'(r_0)}{4\pi} = \frac{g^2 r_0}{2 \pi }-\frac{g^2 m}{2 \pi  \left(g^2 r_0^2+k\right)}
+ {\cal O}(m^2)\,.
\ee
Since entropy formula (\ref{genentropy}) involves only $f'(r_0)$, we may also define the entropy as
\be
S=\fft{\kappa_0 \omega m}{2g^2r_0}  - \fft{\kappa_0 \omega k}{2g^2} + {\cal O}(m^2)\,.
\ee
In fact even with the curvature singularity, it can be verified that the identity (\ref{waldid}) holds, with $(\delta {\cal H})_{r=r_0}=\omega/(4\pi)\delta m$.  There is however a subtlety evaluating $(\delta {\cal H})_{r=r_0}$.  In a usual black hole, as observed in \cite{wald1,wald2}, only the $\delta Q$ term in (\ref{varyH}) contributes to $(\delta {\cal H})_{r=r_0}$, which gives rise to $T\delta S$.  For this solution, both the $\delta Q$ and $i_{\xi} \Theta$ term are divergent on the horizon and they conspire to give a finite $(\delta {\cal H})_{r=r_0}$ that is identical to $(\delta{\cal H})_{r\rightarrow \infty}$.  That the $i_{\xi} \Theta$ term also contributes on the horizon was rare and only seen previously in black holes of Horndeski gravity \cite{Feng:2015oea,Feng:2015wvb}.

It is worth remarking that the small mass black hole has two integration constants, the mass $m$ and the horizon radius $r_0$.  It is rather unusual that these two parameters are independent of each other.  Whilst we have seen that the Wald identity (\ref{waldid}) is indeed satisfied, it is not clear how one can extract a sensible first law.

\subsection{$D=5$ AdS black hole}

In five dimensions with negative cosmological constant, the critical condition is given by
\be
\Lambda_0 = - 4g^2\,,\qquad \lambda= - \fft{1}{24 g^4}\,.
\ee
Correspondingly, we find that the theory admits the following solution
\be
ds^2 = - f dt^2 + \fft{dr^2}{f} + r^2 d\Omega_{3,k}^2\,,\qquad f=g^2 r^2 + k - \mu\,,\label{d5bh}
\ee
where $d\Omega_{3,k}^2$ denotes the metrics of unit $S^3$, $T^3$ and $H^3$ for $k=-1,0,1$ respectively.  The solution describes an asymptotic AdS black hole with the horizon located at $r_0$, with $\mu=g^2r_0^2 + k$.  The temperature and the entropy from (\ref{genentropy}) are given by
\be
T=\ft{g^2}{2\pi} r_0\,,\qquad S=-\ft12 {\kappa_0} \omega r_0^3 \Big(1 + \fft{3k}{g^2 r_0^2}\Big)\,.
\ee
The completion of the first law $dM=TdS$ yields the mass
\be
M=-\fft{3\kappa_0 \omega}{16\pi g^2} \mu^2\,.
\ee
The result is consistent with the Wald formalism discussed in section \ref{wald}.  We can easily establish that
\be
(\delta {\cal H})_{r=r_0}=T \delta S\,,\qquad (\delta {\cal H})_{r\rightarrow \infty}=\delta M\,.
\ee
Note that the positiveness of the mass and entropy requires that $\kappa_0=-1$, which at critical point does not upset the ghost-free condition.

It is intriguing to mention that the exact black hole solution (\ref{d5bh}) exists also in critical Einstein-Gauss-Bonnet gravity \cite{Fan:2016zfs} or more general Lovelock gravities \cite{Crisostomo:2000bb}.  Furthermore, the mass and entropy match with those \cite{Fan:2016zfs} in critical Einstein-Gauss-Bonnet gravity as well.

\subsection{$D=4$ AdS wormbrane}

When the bare cosmological constant $\Lambda_0$ is negative, the critical point in four dimensions is given by (\ref{d4crit}).  The cosmological solution discussed in section \ref{d4bounce}  can be analytically continued to become a static solution with the metric
\be
ds_4^2 = dr^2 + \cosh^2(g\,r) (-dt^2 + dx^2 + dy^2)\,.
\ee
The metric is like a concave domain wall smoothly connecting two asymptotic AdS regions with three-dimensional Minkowski spacetimes as the boundaries.  Such AdS wormbrane solution would necessarily violate the null-energy condition in Einstein theory, but can arise in the pure gravity sector of critical Einsteinian cubic gravity.

\section{Anisotropic dS bounce universes in $D=4,5$}

The AdS black hole solutions constructed in the previous section assumes that the bare cosmological constant $\Lambda_0$ at the critical point is negative.  For $\Lambda_0>0$, these static black holes with $k=0$ become naturally cosmological solutions where the $r$ coordinate becomes time like.  Making appropriate coordinate transformation, we find
\bea
ds_4^2 &=& -dt^2 + \sinh^2(H_0 t) dz^2 + \cosh^2(H_0 t) (dx^2 + dy^2)\,,\label{d4an}\\
ds_5^2 &=& - dt^2 + \sinh^2 (H_0 t) dz^2 + \cosh^2 (H_0 t) (dx^2 + dy^2 + dw^2)\,,\label{d5an}
\eea
where the effective Hubble constant $H_0$ is given by (\ref{d4h0}) and (\ref{d5h0}) for four and five dimensions respectively.  Note that the $D=5$ solutions can be also constructed in critical Einstein-Gauss-Bonnet gravity of\cite{Fan:2016zfs}.  These Bianchi-type IX solutions are anisotropic and outside the FLRW class of cosmology. However, as $t\rightarrow \pm \infty$, the metrics become isotropic de Sitter spacetimes.  At $t=0$, the size of the $z$ direction shrinks to zero, but it is not a curvature singularity.  Instead, the geometry at the region of $t=0$ is a direct product of a two-dimensional Milne universe and Euclidean $\mathbb R^2$ or $\mathbb R^3$.  Thus we obtain exact and smooth anisotropic cosmology solutions that describe bounces between two isotropic dS spacetimes with flat spatial directions.  The anisotropicity between $z$ and the remaining space coordinates becomes insignificant for sufficiently-large e-foldings.  In fact, such anisotropic solutions are not rare, and can be found also in Einstein gravity with a positive cosmological constant in general dimensions:
\bea
ds^2_D &=& -dt^2 + \sinh^2(\ft12(D-1) H_0t) [\cosh(\ft12(D-1)H_0 t)]^{-\fft{D-3}{D-1}}\, dz^2\cr &&+ [\cosh(\ft12(D-1) H_0 t)]^{\fft2{D-1}}\, dx^i dx^i\,.
\eea
It is easy to verify that these metrics are Einstein with $R_{\mu\nu}=(D-1) H_0^2\, g_{\mu\nu}$.

\section{Conclusions}

In this paper we studied Einsteinian cubic gravities at the critical point where the linearized equations of motion on the maximally-symmetric vacuum automatically vanishes.  We showed that cosmological solutions describing isotropic bounce universes could arise in four and five dimensions, but not beyond.  In particular, we obtained an exact bounce universe in four dimensions.  The comoving time runs from minus infinity to plus infinity and the metric interpolates between two de Sitter vacua with flat spatial sections.  However, a linear analysis indicated that there existed a singular scalar mode and hence the solution is not stable.  In five dimensions, we adopted a numerical approach and obtained a bounce universe.  Instead of connecting two de Sitter spacetimes, there exists a curvature singularity before the bounce takes the place.  This suggests that in higher derivative gravities, a bounce in cosmology may not necessarily resolve the initial cosmic singularity.

   We also obtained exact AdS black holes in four and five dimensions in the critical theory with negative cosmological constant.  Exact such solutions were known to exist in critical Lovelock gravities in odd dimensions only; however, critical Einsteinian cubic gravity admits such a solution in four dimensions as well. Furthermore, we also find an AdS wormbrane in four dimensions that smoothly connects two flat AdS boundaries.  When the cosmological constant is positive, the planar black holes can be analytically continued to become smooth anisotropic cosmological solutions that bounce between two isotropic de Sitter spacetimes with flat spatial directions.  The richness of the solutions, both static and cosmological, in critical Einsteinian gravity makes the theory interesting for investigations.

\section*{Acknolwedgement}

The work is supported in part by NSFC grants NO. 11475024, NO. 11175269, and NO. 11235003.

\end{document}